# Time delay of photoionization by Endohedrals

M.Ya. Amusia[1, 2] and L.V. Chernysheva[2]

[1]*Racah Institute of Physics, the Hebrew University, Jerusalem, 91904 Israel*
[2]*Ioffe Physical-Technical Institute, St. Petersburg, 194021 Russia*

**Abstract**

In this Letter, we investigate the time delay of photoelectrons by fullerenes shell in endohedrals. We present general formulas in the frame of the random phase approximation with exchange (RPAE) applied to endohedrals A@$C_N$ that consist of an atom A located inside of a fullerenes shell constructed of N carbon atoms C.

We calculate the time delay of electrons that leave the inner atom A in course of A@$C_N$ photoionization. Our aim is to clarify the role that is played by $C_N$ shell, its static potential, and polarization. As concrete examples of A we have considered Ne, Fr, Kr and Xe, and as fullerene we take $C_{60}$.

The presence of the $C_{60}$ shell manifests itself in powerful oscillations of the time delay $\tau_{nl}(\omega)$ of an electron that is ionized from a given subshell *nl* by a photon with energy $\omega$. Calculations are performed for outer, subvalent and *d*-subshells.

**1.** The aim of this Letter is to clarify how the presence of a fullerenes shell $C_N$ that consist of N carbon atom C modifies time delay of electrons that are ejected from subshell *nl* of an atom A stuffed inside the fullerene forming an endohedral A@$C_N$. Temporal description of physical processes is common in classical physics. Till recently, for quantum mechanical objects dominative was the energy representation. In its frame atomic photoionization is considered as a process of absorbing a given energy photon by an atom or an atom-like object that results in emission of one or several electrons having well-defined energies and directions of their linear momenta. In the frame of such an approach questions like how much time does it take for an electron to live the atom after absorbing a photon or how long does it take for an Auger decay to proceed has no sense.

The temporal description of processes in quantum-mechanical objects was developed long ago and published in seminal papers of L. Eisenbud [1], E. Wigner [2] and F. Smith [3], who introduced the so-called EWS time delay as a quantum dynamical observable. Starting with a simple question "How much time does it take for a particle to penetrate through a potential barrier?" they derived the general expression for the intrinsic time delay due of any physical process. This quantity is a well-defined characteristic only for short-range interaction of particle wave packet with a target, demanding some caution when applied to photoionization of atoms.

EWS formulas remained almost untouched for decades until relatively recently, when short, attosecond pulses were created and applied to photoionization. Then with the help of EWS formulas it became possible to present a temporal picture of atomic and molecular processes. Investigations of times delay in atomic and molecular photoionization is now a rapidly developing field of research (see, for example, [4-9] and references therein).

The fullerenes shell is an additional source of time delay in photoionization of endohedrals. It is essential to have in mind that attosecond spectroscopy of molecular systems provides a remarkable possibility of insight into the process of the outgoing wave packet formation within the molecule, where the multi scattering of the electron wave by the atomic constituents is very important. Of special interest are empty inside fullerenes $C_N$ stuffed with atoms A, called endohedrals A@$C_N$. Among this group of objects relatively easy tractable are noble gas endohedrals formed by noble gases He, Ne, Ar and Xe stuffed inside the almost spherically symmetric $C_{60}$. We will treat these objects in the frame of random phase approximation with exchange (RPAE) used most straightforwardly for closed subshell atoms [10] and generalized to include effects of $C_{60}$ substituting it by a spherical static potential well and including its ability to be dipolar polarized by the incoming photon beam [11].

**2.** As it was demonstrated in [1-3], the time delay of a physical process $\tau$ as a function of its energy $\omega$ [1] is connected to the phase of the amplitude $f$ of the process under consideration by the following relation

$$\tau(\omega)=\frac{d}{d\omega}f(\omega)\equiv \mathrm{Im}\left[\frac{1}{f(\omega)}\frac{df(\omega)}{d\omega}\right] \tag{1}$$

In one-electron Hartree-Fock (HF) approximation to photoionization the matrix element $\left\langle \psi_{\mathbf{k}}^{(-)}\middle|z\middle|\phi_i\right\rangle$ [10, 12] determines the amplitude of a transition of an atomic electron in the state $i$ with the wave-function $\phi_i(\mathbf{r})=R_{n_il_i}(r)Y_{l_im_i}(\hat{\mathbf{r}})$ to the continuous state of an electron with momentum $\mathbf{k}$ after absorbing a linearly polarized photon, in which case the polarization vector is directed along the z axis and $z=\sqrt{4\pi/3}rY_{10}(\hat{\mathbf{r}})$ [2] [10, 12, 5]. Thus, in HF approximation $f(\omega)\Rightarrow f_{n_il_i}^{HF}(\varepsilon,\hat{\mathbf{k}})$ is determined by the following formula:

$$f_{n_il_i}^{HF}(\varepsilon,\hat{\mathbf{k}})\propto\left\langle \psi_{\mathbf{k}}^{(-)}\middle|z\middle|\phi_i\right\rangle=\frac{(2\pi)^{3/2}}{k^{1/2}}\sum_{\substack{l=l_i\pm1\\ m=m_i}} e^{i\delta_l}i^{-l}Y_{lm}(\hat{\mathbf{k}})(-1)^m\begin{pmatrix} l & 1 & l_i\\ -m & 0 & m_i\end{pmatrix}\left\langle \varepsilon l\|r\|n_il_i\right\rangle, \tag{2}$$

where $\omega=\varepsilon+I_{n_il_i}$, $\varepsilon=k^2/2$

Here

$$\left\langle \varepsilon l\|r\|n_il_i\right\rangle=\sqrt{(2l+1)(2l_i+1)}\begin{pmatrix} l & 1 & l_i\\ 0 & 0 & 0\end{pmatrix}\int_0^\infty r^2drR_{\varepsilon l}(r)rR_{n_il_i}(r)=e^{i\pi l_>}\sqrt{l_>}\int_0^\infty r^2drR_{\varepsilon l}(r)rR_{n_il_i}(r), \tag{3}$$

where $l_>$ denotes the larger value of angular momentum, $l$ or $l_i$. The continuous spectrum radial wave function are normalized in the energy scale according to the relation $\left\langle \varepsilon l\middle|\varepsilon' l\right\rangle=\delta(\varepsilon-\varepsilon')$ and their asymptotic is given by the following formula [13]

$$R_{\varepsilon l}(r)\Big|_{r\to\infty}=\sqrt{\frac{2}{\pi k}}\frac{1}{r}\sin\left(kr-\frac{l\pi}{2}+\delta_l\right). \tag{4}$$

---

[1] Atomic system of units $e=m=\hbar=1$ is used in this paper, where $e$ is the electron charge, $m$ is its mass and $\hbar$ is Planck's constant.

[2] A roof sign above a vector as e.g. in $\hat{\mathbf{r}}$ means a unit vector in the direction of vector $\mathbf{r}$.

Here $\delta_l \equiv \delta_l(\varepsilon)$ is the continuous spectrum electron scattering phase in HF approximation [10-12].

Usually, in attosecond time delay measurements the forward direction amplitude is of interest. In this case one has to substitute $Y_{lm}(\hat{\mathbf{k}})$ by $\sqrt{(2l+1)/4\pi\delta_{m0}}$. As is seen from (1), time delay is not affected by real numeric factors in the total amplitude since they do not affect the phase shift. This is why we can define the photoionization amplitude in forward direction omitting numeric factors from (2). Thus, for the aims of time delay determination we have in HF:

$$
\begin{aligned}
f_{n_il_i}^{HF}(\varepsilon,0) &= \frac{1}{\sqrt{k(2l_i+1)}}\sum_{l=l_i\pm1} e^{i[\delta_l(\varepsilon)+l_>\pi-l\pi/2]}\sqrt{l_>}\left\langle \varepsilon l\left\|r\right\|n_il_i\right\rangle, \\
\operatorname{Im} f_{n_il_i}^{HF}(\varepsilon,0) &= \frac{1}{\sqrt{k(2l_i+1)}}\sum_{l=l_i\pm1} \sin\left[\delta_l(\varepsilon)+l_>\pi-l\pi/2\right]\sqrt{l_>}\left\langle \varepsilon l\left\|r\right\|n_il_i\right\rangle,. \\
\operatorname{Re} f_{n_il_i}^{HF}(\varepsilon,0) &= \frac{1}{\sqrt{k(2l_i+1)}}\sum_{l=l_i\pm1} \cos\left[\delta_l(\varepsilon)+l_>\pi-l\pi/2\right]\sqrt{l_>}\left\langle \varepsilon l\left\|r\right\|n_il_i\right\rangle
\end{aligned}
\tag{5}
$$

Time-delay in HF is given, according to (1) and (5)

$$
\tau_{n_il_i}^{HF}(\varepsilon) = \frac{d}{d\varepsilon}\varphi_{n_il_i}^{HF}(\varepsilon) = \frac{d}{d\varepsilon}\arctan\frac{\operatorname{Im} f_{n_il}^{HF}(\varepsilon)}{\operatorname{Re} f_{n_il}^{HF}(\varepsilon)}. \tag{6}
$$

Taking into account the RPAE electron correlations leads to RPAE amplitude $f_{n_il_i}^{RPAE}(\varepsilon,0)$ that is obtained from (5) after substituting $\left\langle \varepsilon l\left\|r\right\|n_il_i\right\rangle$ by its RPAE value $\left\langle \varepsilon l\left\|D(\omega)\right\|n_il_i\right\rangle$, which is determined by equation [10, 11]

$$
\begin{aligned}
\left\langle \varepsilon l\left\|D(\omega)\right\|n_il_i\right\rangle &= \left\langle \varepsilon l\left\|r\right\|n_il_i\right\rangle + \frac{1}{3}\sum_{\substack{n'l'\\ n_jl_j}}\int_0^\infty d\varepsilon'\left[\frac{\left\langle \varepsilon'l'\left\|D(\omega)\right\|n_jl_j\right\rangle\left\langle n_jl_j,\varepsilon l\left\|U\right\|\varepsilon'l',n_il_i\right\rangle}{\omega-\varepsilon'+\varepsilon_{n_il_i}+i\eta}\right. \\
&\left.+\frac{\left\langle n_jl_j\left\|D(\omega)\right\|\varepsilon'l'\right\rangle\left\langle \varepsilon'l',\varepsilon l\left\|U\right\|n_jl_j,n_il_i\right\rangle}{\omega+\varepsilon'-\varepsilon_{n_il_i}}\right], \quad \eta\to+0
\end{aligned}
\tag{7}
$$

Here summation over $n_jl_j$ includes all occupied electron subshell of an atom, while summation over $n'l'$ includes discrete vacant excited states. Here matrix elements of $U$ are combinations of direct and exchange Coulomb interelectron interaction $V = 1/|\mathbf{r}_1-\mathbf{r}_2|$. More details on solving (6) and corresponding computing programs can be found in [12].

Additional phase $\varphi_{D_l}(\varepsilon)$ to the photoionization amplitude in RPAE comes from the term in (7) that contains $\eta\to+0$. The matrix element $\left\langle \varepsilon l\left\|D(\omega)\right\|n_il_i\right\rangle$ can be presented as

$$\left\langle \varepsilon l \left\| D(\omega) \right\| n_i l_i \right\rangle = \mathrm{Re}\left\langle \varepsilon l \left\| D(\omega) \right\| n_i l_i \right\rangle + i\,\mathrm{Im}\left\langle \varepsilon l \left\| D(\omega) \right\| n_i l_i \right\rangle \equiv \left\langle \varepsilon l \left\| \tilde{D}(\omega) \right\| n_i l_i \right\rangle e^{i\varphi_{D_l}(\varepsilon)}, \qquad .(8)$$

where

$$\varphi_{D_l}(\varepsilon) = \arctan\left[ \mathrm{Im}\left\langle \varepsilon l \left\| D(\omega) \right\| n_i l_i \right\rangle / \mathrm{Re}\left\langle \varepsilon l \left\| D(\omega) \right\| n_i l_i \right\rangle \right]. \qquad (9)$$

As a result, the following relation defines the RPAE amplitude of an isolated atom

$$f_{n_i l_i}^{RPAE}(\varepsilon, 0) = \frac{1}{\sqrt{k(2l_i+1)}} \sum_{l=l_i \pm 1} \sqrt{l_>} \exp\left\{ i\left[ \delta_l(\varepsilon) + \varphi_{D_l}(\varepsilon) + l_> \pi - l\pi/2 \right] \right\} \left\langle \varepsilon l \left\| \tilde{D}(\omega) \right\| n_i l_i \right\rangle. \qquad (10)$$

Time-delay in RPAE is given, similar to (6), by

$$\tau_{n_i l_i}^{RPAE}(\varepsilon) = \frac{d}{d\varepsilon} \varphi_{n_i l_i}^{RPAE}(\varepsilon) = \frac{d}{d\varepsilon} \arctan \frac{\mathrm{Im}\, f_{n_i l}^{RPAE}(\varepsilon)}{\mathrm{Re}\, f_{n_i l}^{RPAE}(\varepsilon)}. \qquad (11)$$

The determination of the phase shift via $\arctan$ leads to some uncertainties. Therefore in order to perform numeric calculations it is more efficient to deal directly with time delay, i.e. with the energy derivative of the ratio of imaginary and real parts of the amplitude. This can be achieved using the following formula:

$$\tau_{n_i l_i}^{RPAE}(\varepsilon) = \frac{d}{d\varepsilon} \arg f_{n_i l_i}^{RPAE}(\varepsilon) = \frac{\left[ d\,\mathrm{Im}\, f_{n_i l_i}^{RPAE}(\varepsilon)/d\varepsilon \right] \mathrm{Re}\, f_{n_i l_i}^{RPAE}(\varepsilon) - \left[ d\,\mathrm{Re}\, f_{n_i l_i}^{RPAE}(\varepsilon)/d\varepsilon \right] \mathrm{Im}\, f_{n_i l_i}^{RPAE}(\varepsilon)}{\left[ \mathrm{Re}\, f_{n_i l_i}^{RPAE}(\varepsilon)^2 + \mathrm{Im}\, f_{n_i l_i}^{RPAE}(\varepsilon)^2 \right]}. \qquad (12)$$

**3.** In principle, it is possible to directly apply equation (7) to endohedrals A@$C_{60}$. To do this, one has to include in summation over $n_j l_j$ all one-electron occupied levels in the endohedral and extend summation over $n'l'$ including all discrete vacant excited states. This is almost impossible and unnecessary. Instead, we substitute $C_{60}$ by a static potential $W(r)$ that corresponds to reasonable distribution of electric charges and satisfactorily reproduces the experimental value of an s-electron affinity of $C_{60}^-$ that is equal to $E_s = -2.65eV$ [14]. In this approach, called $RPAE_C$ the electrostatic potential of the fullerene shell as a whole is a sum of the positive potential of carbon atom nuclei smeared over a sphere with radius $R$, and a negative potential created by the electron clouds. The potential $W(r)$ is added to the HF atomic A potential. It affects $HF_C$ and $RPAE_C$ matrix elements and phases [11, 12]. The $C_{60}$ potential has Lorentz-shape bubble potential, as suggested in [15]

$$W(r) = -W_{\max} \frac{d^2}{(r-R)^2 + d^2}. \qquad (13)$$

Here $R$ is the fullerenes radius. This potential, contrary to often used square-well potentials [7, 16], belong to the class of bottom-curved potentials that reproduce physically satisfactory the electrical charge distribution of $C_{60}$ shell [15].

In addition to static potential $W(r)$ it is essential to take into account the dipole polarization of the fullerene shell that essentially modifies the action of the incoming light beam upon the ionizing inner atom A. Assuming for simplicity that the atomic radius $R_a$ is much smaller than $R$, one can in the RPAE frame express the effect of dipole polarization factor $G_{C_N}(\omega)$ that connects the endohedral photoionization amplitude in "atomic" RPAE $f_{n_il_i}^{RPAE_C}(\varepsilon)$ with the endohedral RPAE photoionization amplitude $f_{n_il_i}^{A@C_N}(\varepsilon)$ by a simple relation $f_{n_il_i}^{A@C_N}(\varepsilon) = G_{C_N}(\omega) f_{n_il_i}^{RPAE_{C_N}}(\varepsilon)$. The polarization factor is expressed via dipole polarizability of $C_{60}$ [11, 12, 14]:

$$G_{C_N}(\omega) = \left[1 - \frac{\alpha_C^d(\omega)}{R^3}\right] \equiv \tilde{G}_{C_N}(\omega) e^{i\varphi_C(\omega)}, \tag{14}$$

where $\alpha_C^d(\omega)$ is the $C_{60}$ dipole dynamic polarizability, determined by the total photoionization cross section $\sigma(\omega)$ of $C_N$ (see, e.g.[11]):

$$\mathrm{Re}\,\alpha_C^d(\omega) \cong \frac{c}{2\pi^2} \int_{I_C}^{\infty} \frac{\sigma(\omega')d\omega'}{\omega'^2 - \omega^2};\ \mathrm{Im}\,\alpha_C^d(\omega) = \frac{c\sigma(\omega)}{4\pi\omega};\ \frac{c}{2\pi^2}\int \sigma(\omega)d\omega \cong N_e\ . \tag{15}$$

Here $c$ is the speed of light and $N_e$ is the total number of electrons in $C_N$. Similar to (5) and (10), the following relations determines the zero angle photoionization amplitude of A@$C_N$

$$f_{n_il_i}^{A@C}(\varepsilon,0) = \frac{\tilde{G}_{C_N}(\omega)}{\sqrt{k(2l_i+1)}} \sum_{l=l_i\pm1} \sqrt{l_>} \exp\left\{ i\left[\delta_l(\varepsilon) + \varphi_{D_l}(\varepsilon) + \varphi_C(\omega) + l_{>\pi} - l\pi/2\right]\right\} \left\langle \varepsilon l \left\| \tilde{D}(\omega) \right\| n_i l_i \right\rangle, \tag{16}$$

and the time delay in photoionization of A@$C_N$

$$\tau_{n_il_i}^{A@C_N}(\varepsilon) = \frac{\left[d\,\mathrm{Im} f_{n_il_i}^{A@C_N}(\varepsilon)/d\varepsilon\right]\mathrm{Re} f_{n_il_i}^{A@C_N}(\varepsilon) - \left[d\,\mathrm{Re} f_{n_il_i}^{A@C_N}(\varepsilon)/d\varepsilon\right]\mathrm{Im} f_{n_il_i}^{A@C_N}(\varepsilon)}{\left[\mathrm{Re} f_{n_il_i}^{A@C_N}(\varepsilon)^2 + \mathrm{Im} f_{n_il_i}^{A@C_N}(\varepsilon)^2\right]}. \tag{17}$$

It is easy and straightforward to generalize the expressions of the amplitude and time delays to non-zero photoelectron emission angle $\theta$. Averaging the formulas for HF, RPAE and A@$C_{60}$ amplitudes over the incoming photon beam direction, one can demonstrate that the generalization to $\theta \neq 0$ can be achieved by adding into the sums over $l$ in the formulas (5), (10), (16) Legendre polynomials $P_l(\cos\theta)$ as multipliers. This will add interference type oscillations into the dependence of $\tau_{n_il_i}(\varepsilon,\theta)$ upon angle at a given photoelectron energy $\varepsilon$.

**4.** We have performed concrete calculations for noble gas endohedrals Ne, Ar, Kr, Xe@$C_{60}$. Parameters of potential (13) are the same as in Ref. [15, 14]. The use of expressions for different contributions $\varphi_{D_l}(\varepsilon)$ and $\varphi_C(\omega)$ to the total phase shifts of the photoionization amplitude leads to technical difficulties, since the phase $\varphi_{\chi}(x) = \arctan[\operatorname{Im}\chi(x)/\operatorname{Re}\chi(x)]$ of a complex function $\chi(x)$ by definition jumps from $-\pi/2$ to $+\pi/2$ at the point $x = a$, when $\operatorname{Re}\chi(x)$ changes sign while going via zero as $\operatorname{Re}\chi(x) = A(x-a)$ (for A, $\operatorname{Im}\chi(x)>0$). The same problem exists with factor $G_{C_N}(\omega)$ since at some value of $\omega$ $\operatorname{Re}G_{C_N}(\omega)$ changes its sign going via zero. It is essential to note that while the phases $\varphi_{D_l}(\varepsilon)$ and $\varphi_C(\omega)$ have jumps in their value, their derivatives, and time delays are not jumping. This follows from numeric calculations and an analytic model that reproduces the $\operatorname{Re}f(\omega) = f'(\Omega)(\omega-\Omega)$ behavior near $\operatorname{Re}f(\Omega)=0$ point.

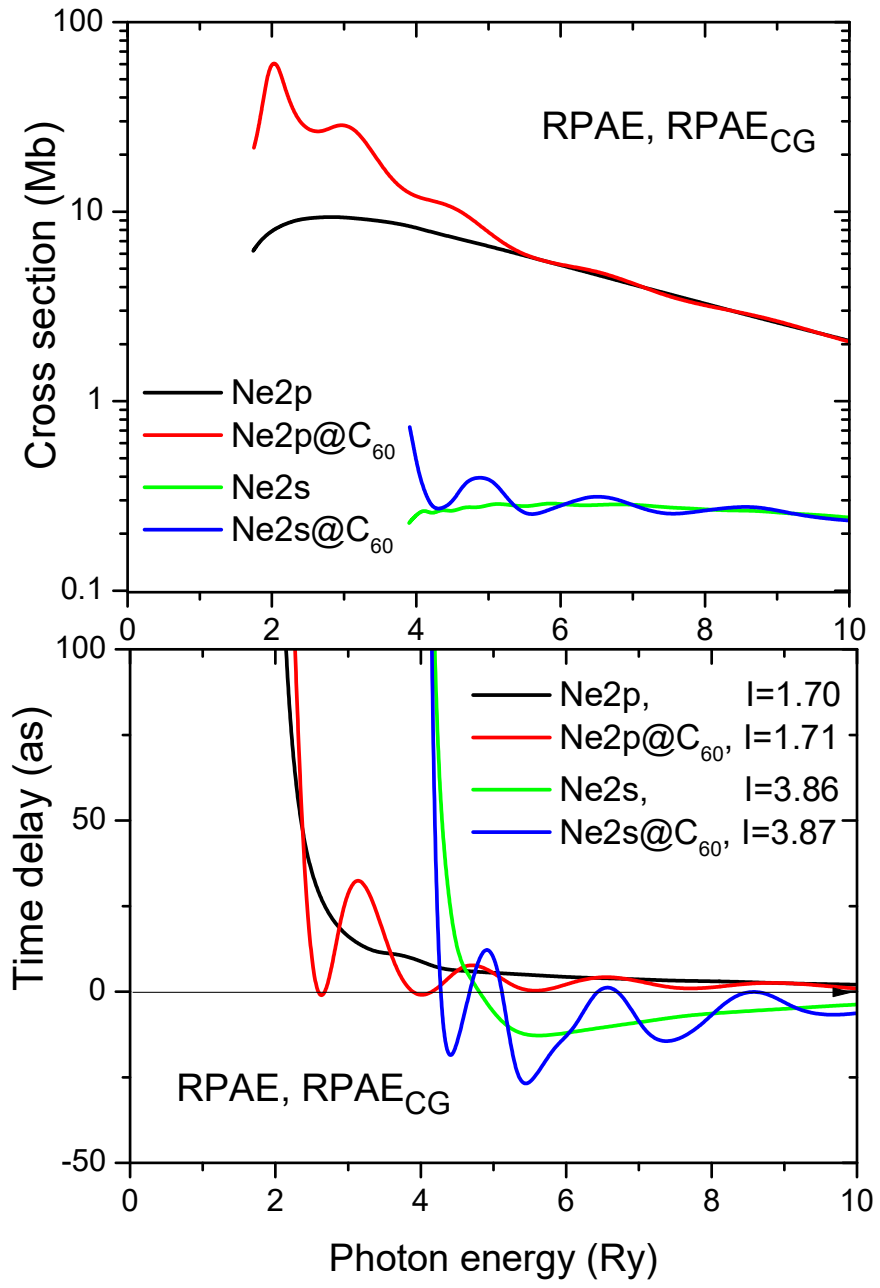

Fig. 1. Partial photoionization cross sections and time delays for 2p and 2s subshells of Ne and Ne@$C_{60}$

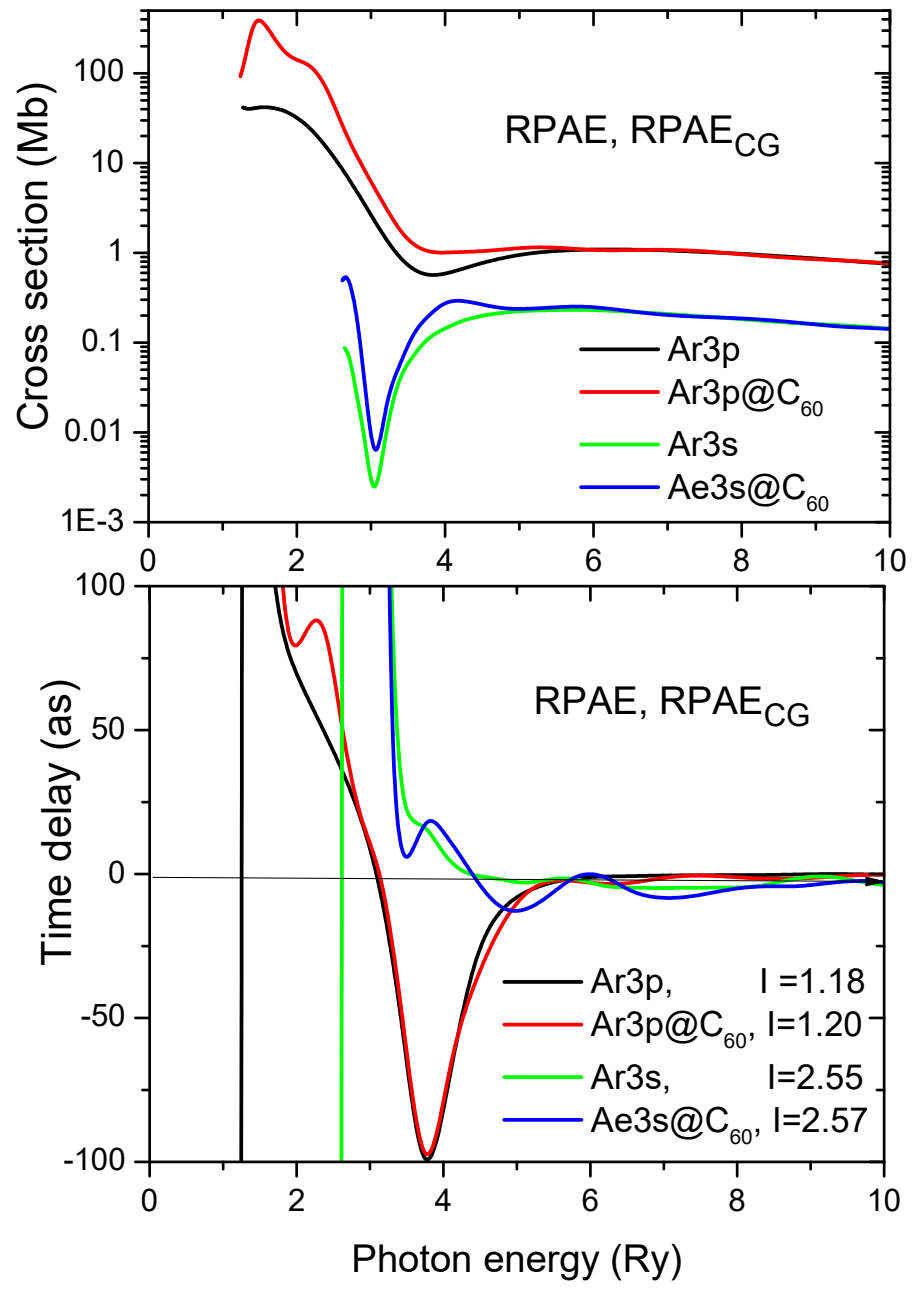

Fig. 2. Partial photoionization cross sections and time delays for 3p and 3s subshells of Ar and Ar@$C_{60}$.

Determination of $\delta_l(\varepsilon)$ is not simple because to this phase contributes not only a short-range HF potential but also a long-range Coulomb tail. This reflects the fact that photoelectron feels the attractive field of single-charged endohedral ion. It is known that Eq. (4) is incorrect in the presence of a Coulomb field and along with *kr* in the argument of the sin function, has also a logarithmic term $k^{-1}\ln 2kr$. The corresponding peculiarities for sufficiently big values of *r* are important only for very small linear momentum *k*. As demonstrated by trial calculations, it took place for $k \le 0.15$. It corresponds to initial continuous spectrum energy $\varepsilon = 0.025$.that is for us sufficiently close to $\varepsilon = 0$.

We calculate photoionization amplitudes and partial cross-section for outer *p*- and subvalent *s*-subshells of noble gases as well as for 3*d*-subshell of Kr and 4d-subshell of Xe. The results of calculations are presented in Fig. 1-4, where cross sections are in Megabarns (Mb, 1Mb=$10^{-18}$cm$^2$), photon energy is in Rydbergs (Ry, 1Ry=13.6 eV), time delay is in Attoseconds (as, 1as=$10^{-18}$sec). For atoms calculations are performed in RPAE and for endohedrals A@$C_{60}$ in RPAE$_C$ with inclusion of G$_C$ that is denoted RPAE$_{CG}$. We are not presenting here results of calculations in HF, since it is well known that for outer, subvalent and *d*-subshells the role of RPAE corrections is big.

In order to better understand the origin of variations of the time delays, we present them in the figures under respective photoionization cross sections. It is known that RPAE satisfactorily describes the partial photoionization cross sections of atoms, particularly noble gas atoms [11]. Thus, comparison between time delay and cross-sections in RPAE reflects almost one-to-one comparison of RPAE time delays dependence on photon energy with that of experimental cross-sections. We see that by using (12) and (17) time delays became reasonably looking smooth functions of $\omega$. The results for Ne in Fig. 1 agree well with results of Fig. 2 in [5]. The inclusion of the $C_{60}$ shell adds to cross sections and time delays prominent oscillations connected to interference caused by crossing the potential barrier $W(r)$. Fig. 1 presents the results of strong time delay oscillations of Ne that are results of $W(r)$ (see (13)) action. The polarization factoк affects mainly the outer subshell, and leads to rather smooth addition to the time delay of all but Ne atom and endohedral.

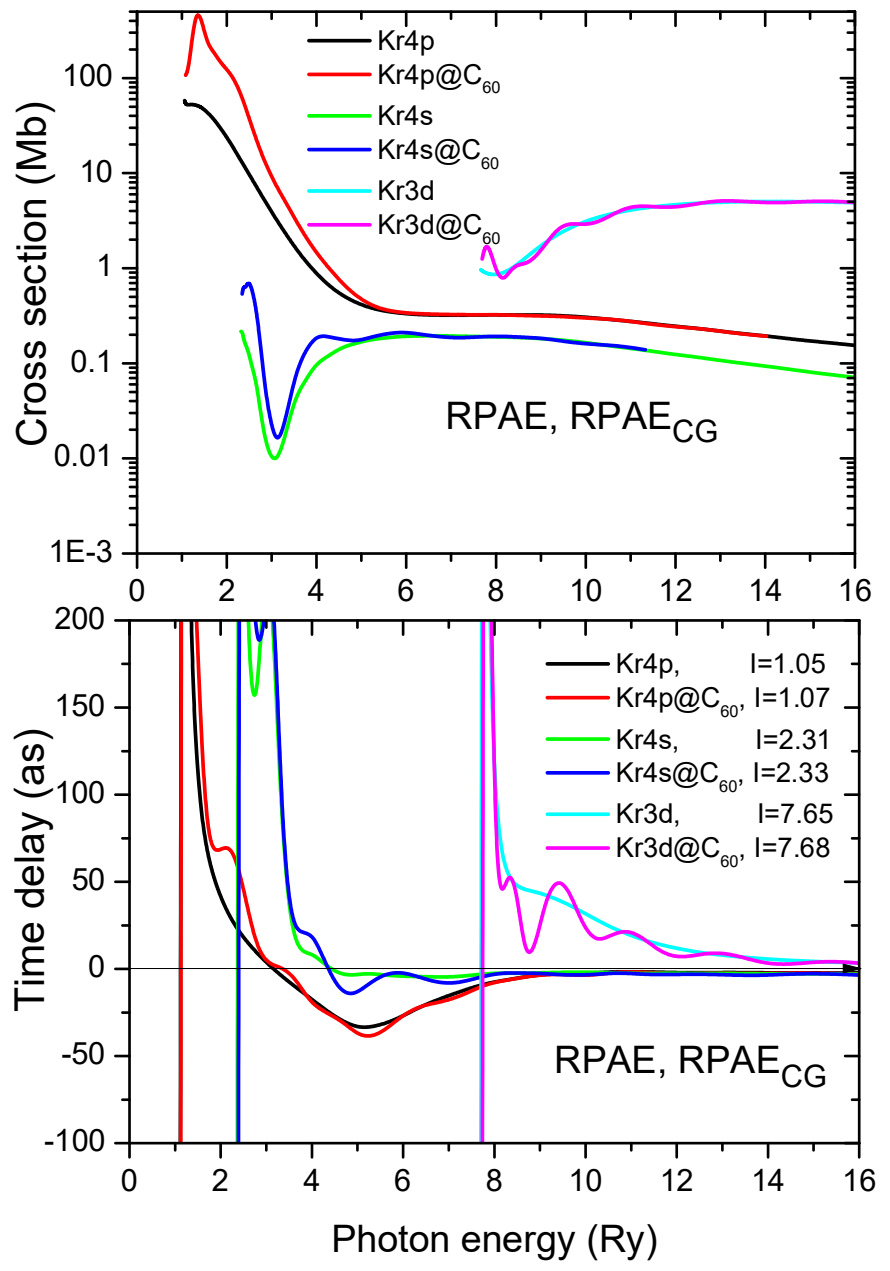

Fig. 3. Partial photoionization cross sections and time delays for 4p, 4s and 3d-subshell of Kr and Kr@$C_{60}$.

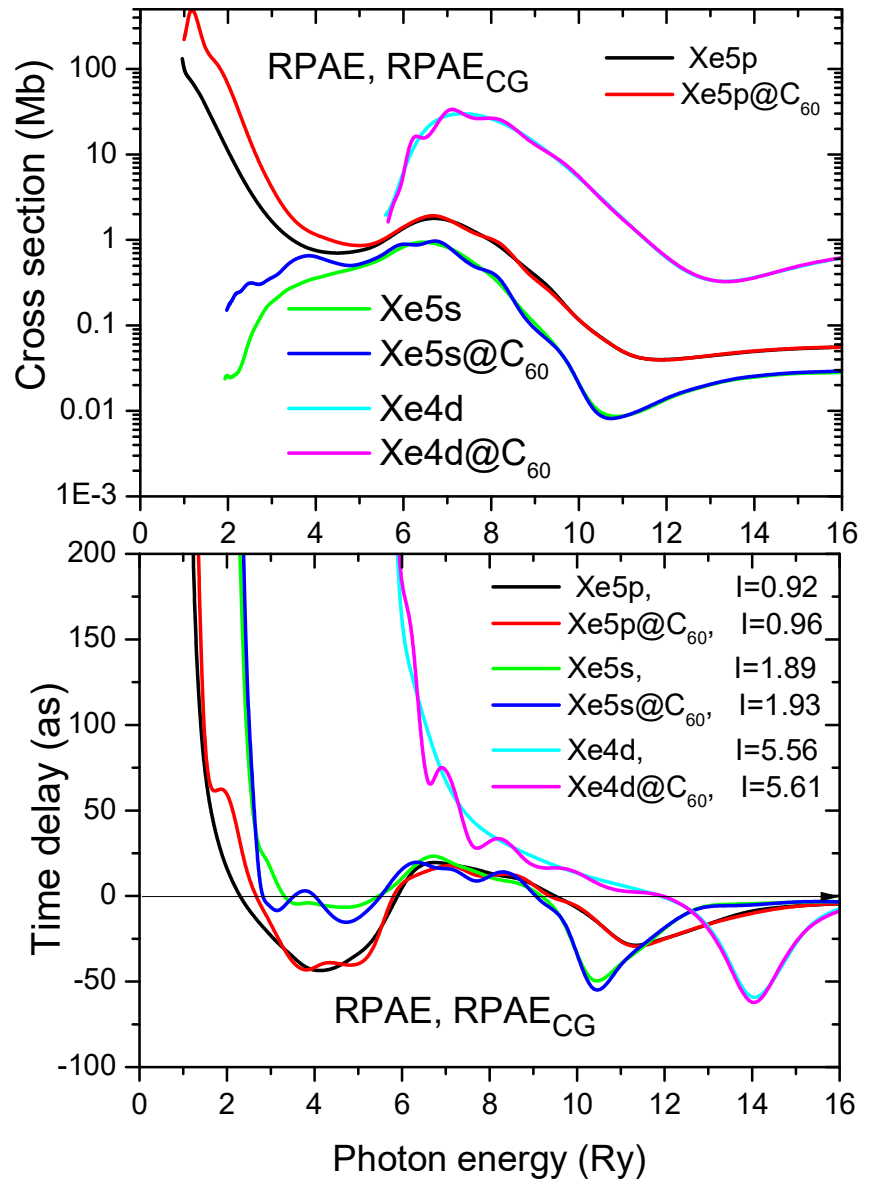

Fig. 4. Partial photoionization cross sections and time delays for 5p, 5s and 5d-subshell of Xe and Xe@$C_{60}$.

It is of interest to note that for Ne the oscillations are particularly strong in the negative time delay direction. Note that large negative delays are in apparent contradiction to the causality principle. From the classical viewpoint, a large negative time delay means that the electron is ejected before the incoming photon acts on the atom or endohedral. However, not too big negative values are compatible with causality due to quantum nature of the photoionization process (see [2]). Fig. 2 depicts the results for Ar atom and respective endohedral. Note a very deep minimum in $\tau_{3p}$ that corresponds to an instant variation not of $\sigma_{3p}(\omega)$, but of $d\sigma_{3p}(\omega)d\omega$. As to the deep minimum in $\sigma_{3s}(\omega)$, it corresponds to a rapid growth in $\tau_{3s}$ with $\omega$ decrease while approaching $3s$ ionization threshold. The presence of $C_{60}$ leads to noticeable variations in photoionization and in time delays. Polarization factor affects time delay weaker than the cross sections.

Fig. 3 and 4 depicts results for Kr and Xe. Maxima in $\tau_{4s,3d}$ for Kr correspond to minims in $\sigma_{4s,3d}$. Times delays $\tau_{4p,4s,3d}$ have powerful maxima at threshold that are affected prominently by polarization factor. Atomic Xe and Xe@$C_{60}$ are of special interest due to presence of the Giant resonance in $4d$-subshell (see, e.g. [10]). We note in Fig. 4 the reflection of this powerful resonance in $\sigma_{5p}(\omega)$ and $\sigma_{5s}(\omega)$ at $\omega \approx 5 \div 9Ry$. It is remarkable, how minima at about 5, 11, 12 and 13$Ry$ in partial photoionization cross-sections are matched with minima in time delays of electrons that are ionized from corresponding subshells. Note that these minima correspond to negative time delays.

**5.** This is the first case when time delays in endohedrals are calculated taking into account the polarization factor and choosing instead of a square well a physically reasonable endohedral potential (13). We believe that by using direct expressions for the time delay (12) and (17) thus escaping calculations of scattering phases as an intermediate step, we have considerably improved the accuracy of our numerical procedure.

The absolute values of the amplitude, together with its phases, totally characterize such a quantum mechanical process as photoionization. Till recently, for dipole photoionization of $n_i l_i$ subshell it was possible to obtain from experiment only the phase differences $[\delta_{l_i+1}(\varepsilon)+\varphi_{D_{l_i+1}}(\varepsilon)-\delta_{l_i-1}(\varepsilon)-\varphi_{D_{l_i-1}}(\varepsilon)]$.Time delay depend upon phases itself, thus supplying in principle new information about the photoionization amplitude. However, to measure the EWS time delay, one have to get rid of the additional contributions that are introduced by the technique of experimental detection of time delay (see e.g. [17] and references therein).

Dealing with attosecond pulses, one must have in mind the limitations that are introduced by the time uncertainty principle. Indeed, e.g. 10 *as* corresponds to energy uncertainty of about 60 eV. It means that instead of a single electron that leaves the target atom or endohedral in the process of ionization of outer or subvalent subshells, more complex target excitations participate in formation of the outgoing wave packet.

In spite of mentioned difficulties, the theoretical investigation of time delays and their experimental measurement is a promising direction in the domain of investigations of light beams interaction with atoms and endohedrals.

**Acknowledgements**

The authors are grateful to Prof. V. Dolmatov who called their attention to the alternative expression for the EWS time delay, presented by (12) and (17).

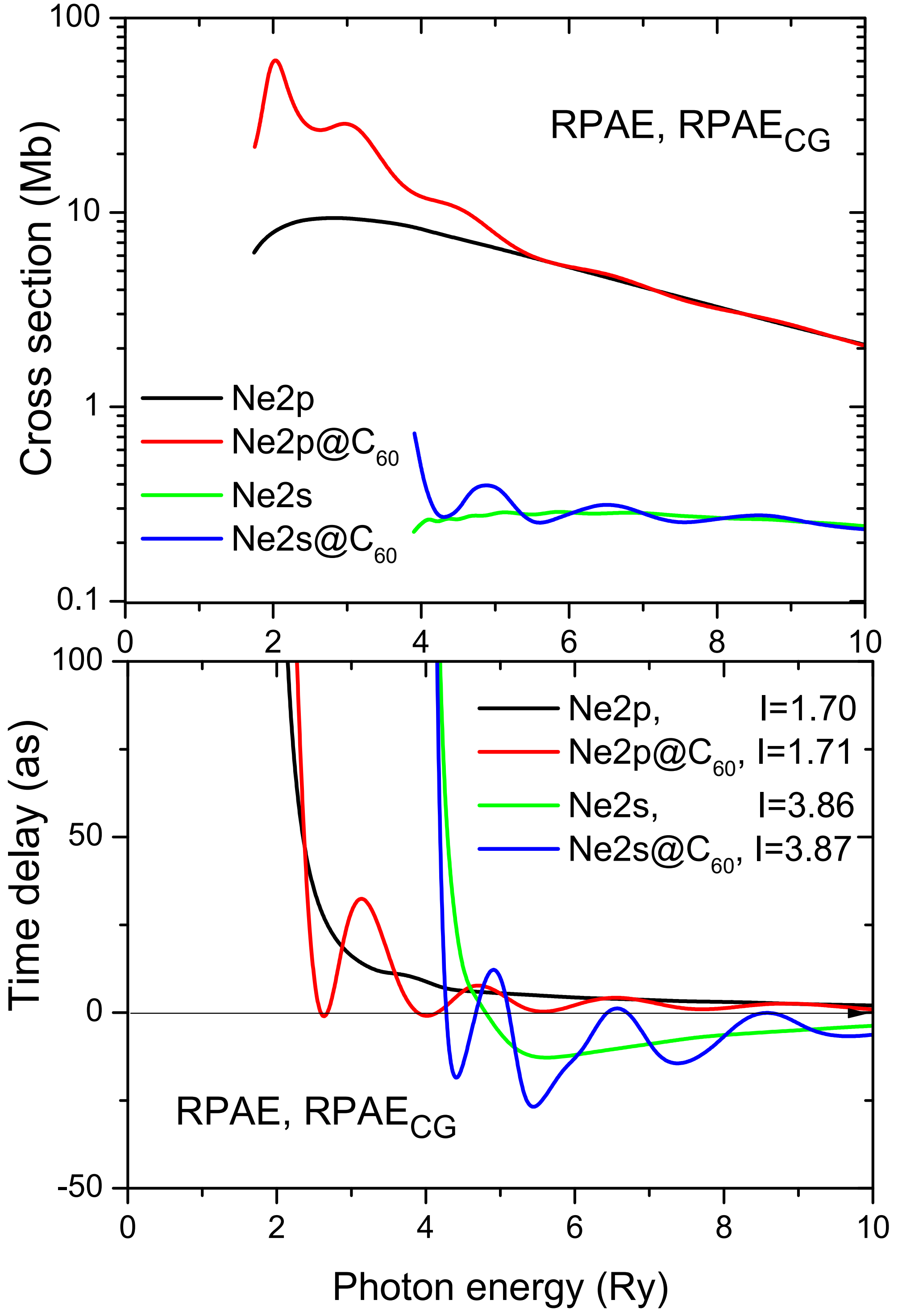

Fig. 1. Partial photoionization cross sections and time delays for 2p and 2s subshells of Ne and Ne@$C_{60}$

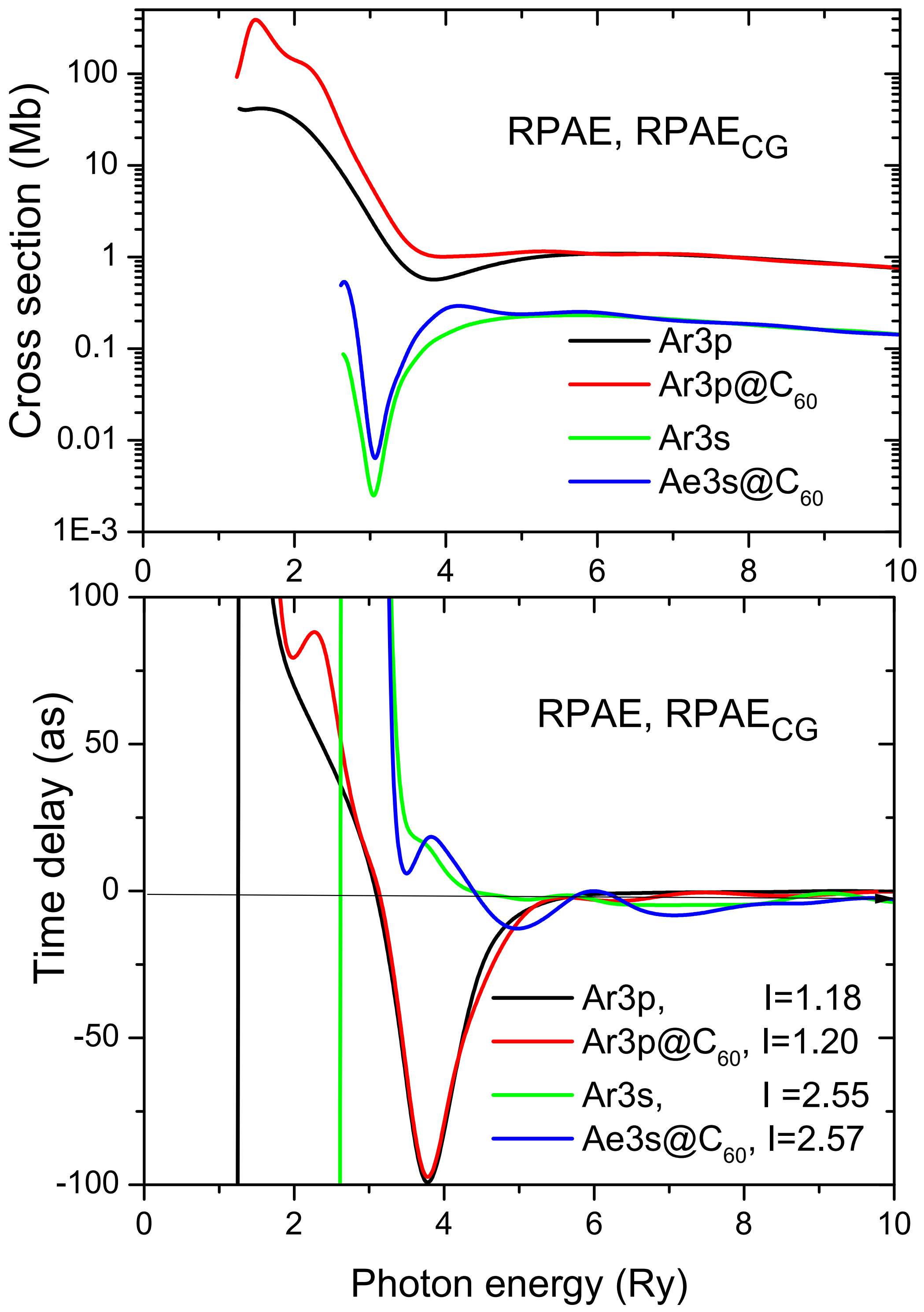

Fig. 2. Partial photoionization cross sections and time delays for 3p and 3s subshells of Ar and Ar@$C_{60}$.

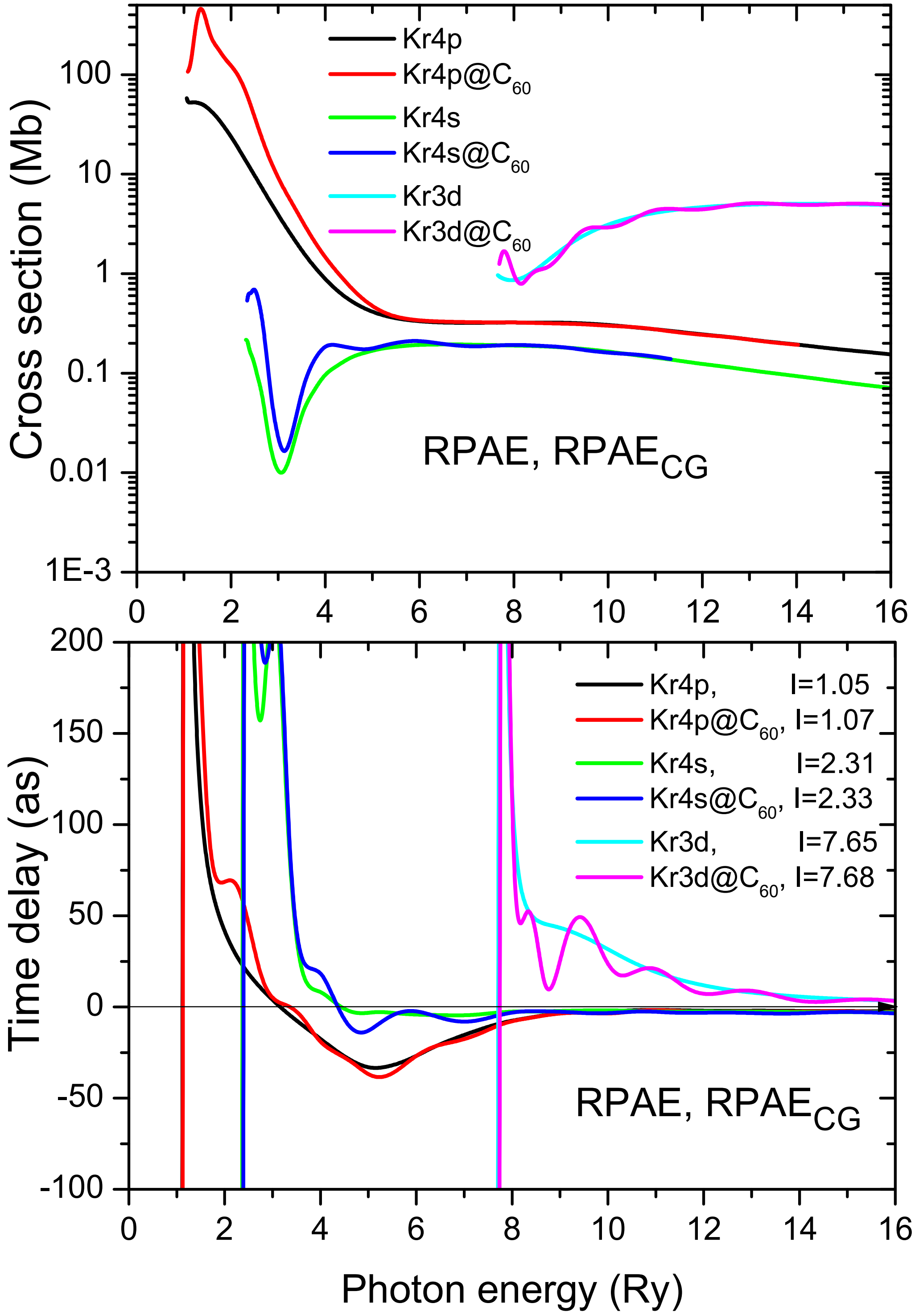

Fig. 3. Partial photoionization cross sections and time delays for 4p, 4s and 3d-subshell of Kr and Kr@$C_{60}$.

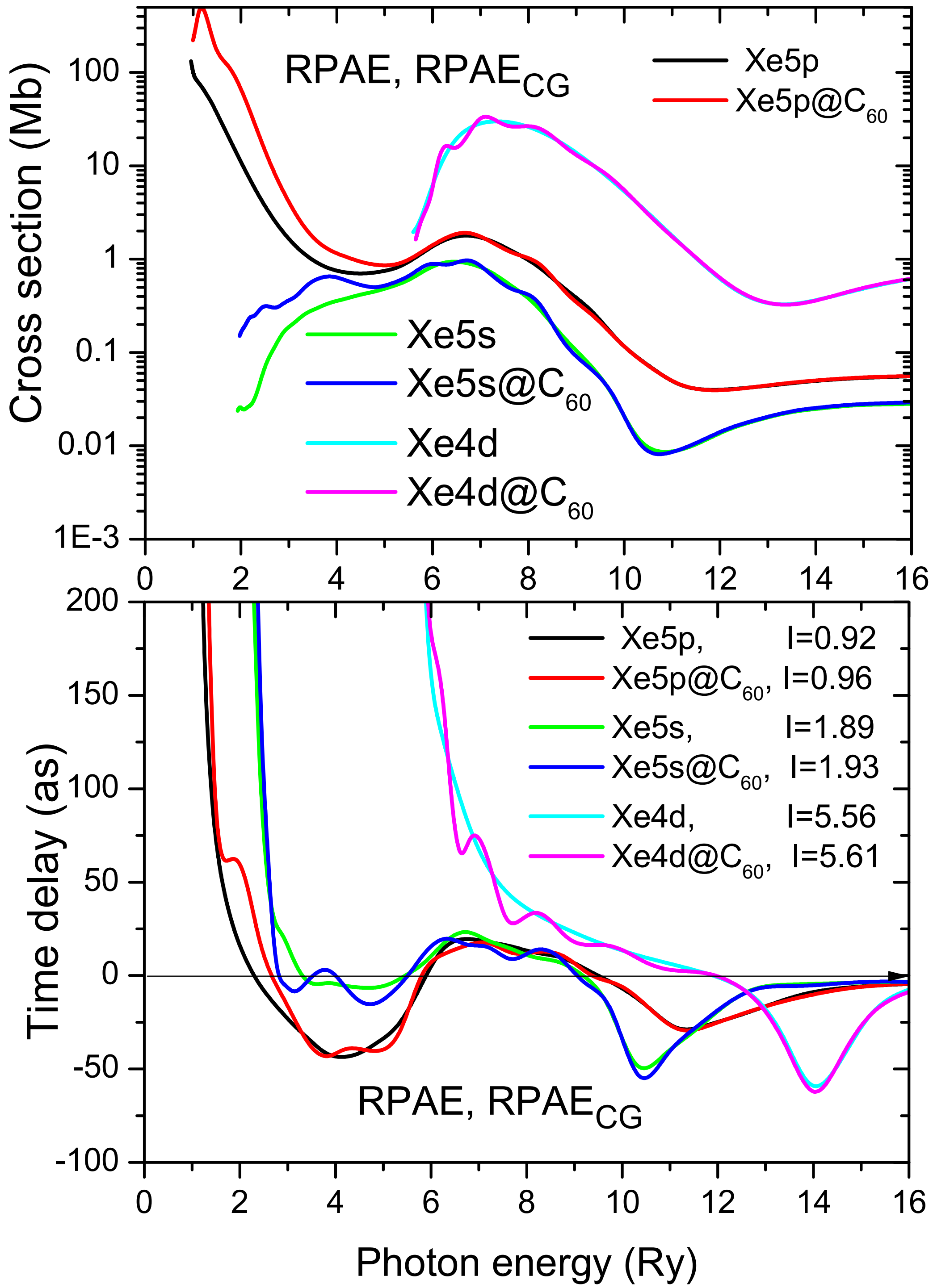

Fig. 4. Partial photoionization cross sections and time delays for 5p, 5s and 5d-subshell of Xe and Xe@$C_{60}$.